\documentclass[aps,prb,twocolumn,oneside,floatfix,sort&compress,
superscriptaddress,showpacs]{revtex4-1} 
\usepackage{graphicx,xcolor}
\usepackage[reqno]{amsmath}
\usepackage[utf8]{inputenc}

\begin{document}
\title{Adaptive Lanczos-vector method for dynamic properties within the
 DMRG}

\author{P. E. Dargel}
\affiliation{Institut f\"ur Theoretische Physik, Georg-August-Universit\"at
G\"ottingen,
37077 G\"ottingen, Germany}
\author{A. Honecker}
\affiliation{Institut f\"ur Theoretische Physik, Georg-August-Universit\"at
G\"ottingen,
37077 G\"ottingen, Germany}
\author{R. Peters}
\affiliation{Department of Physics, Kyoto University, Kyoto 606-8502, Japan}
\author{R.\ M.\ Noack}
\affiliation{Fachbereich Physik, Philipps-Universit\"at Marburg, 35032 Marburg,
Germany}
\author{T. Pruschke}
\affiliation{Institut f\"ur Theoretische Physik, Georg-August-Universit\"at
G\"ottingen,
37077 G\"ottingen, Germany}
                
\date{April 7, 2011}

\begin{abstract}
\pacs{71.10.Pm, 71.10.Fd, 78.20.Bh}
Current widely-used approaches to calculate spectral functions using
the density-matrix renormalization group
in frequency space either necessarily
include an artificial broadening (correction-vector method), have limited
resolution (time-domain density-matrix renormalization group
with Fourier transform method), or are limited to low-energy properties
or single dominant modes (original continued fraction method).  Here we propose
an adaptive Lanczos-vector
method to calculate the coefficients of a continued fraction expansion of the
spectral function iteratively. We show that one can obtain a very accurate
representation of the spectral function very efficiently, and that one can also
directly extract the spectral weights and poles for the discrete system. 
\end{abstract}
\maketitle
Dynamical quantities such as the local density of states, the single-particle
spectral weight, or the dynamical spin or charge correlation functions are
of central importance in theoretical and experimental condensed matter physics.
Since electrons in solids are interacting quantum objects, a reliable
calculation of their properties usually
has to resort to numerical approaches. The density-matrix
renormalization group (DMRG)\cite{*White1992,*White1993} is one such algorithm.
Within the DMRG, the calculation of dynamical quantities is a
considerable challenge. A first attempt by Hallberg was based on
a continued fraction expansion (CFE).\cite{Hallberg1995} Subsequently,
K\"uhner and White showed that this approach is suitable, ``if only the
low-energy part of the correlation function is of interest, or if the bulk of
the weight is in one single peak'',\cite{Kuhner1999} and
applied the correction vector
method,\cite{Kuhner1999,Ramasesha1996}
which since then has successfully been applied to many
model systems.
\cite{Benthien2004,*Shirakawa2009,*Nishimoto2004,*Weichselbaum2009,Raas2005}
However, this approach has the drawback that
one needs to introduce an artificial broadening into the spectra, which can
be viewed as convolution of the true spectral function with a Lorentzian of
width $\eta$. As one is eventually interested in the limit $\eta\to0$, one has
to ``deconvolute'' the spectrum at the end of the calculation. This is,
like analytic continuation of Monte-Carlo data, a numerically
ill-defined procedure.\cite{Raas2005} Furthermore, the calculation of
the correction vector at every step of the DMRG is very time-consuming and
the frequencies that one can address are limited. Another popular approach to
calculating spectral functions is to Fourier-transform time-dependent DMRG
data.\cite{Pereira2009}
To obtain good frequency resolution, one has to calculate 
time-domain data over a long time interval. However, accessing long
time scales is
limited by either a loss of accuracy due to the approximate
nature of the DMRG, or by finite-size effects such as reflections from open
ends. A method that can resolve
spectral features with high resolution and no artificial broadening is
therefore desirable.
Here we present an adaptive Lanczos-vector method (ALM) that takes
advantage of the CFE used in Ref.\ \onlinecite{Hallberg1995}, but
gains efficiency by adapting the basis as in adaptive time evolution.

The spectral function for the operators $\hat{A}$ and $\hat{B}$ is
$2\pi\mathrm{i} \,\rho_{\hat{A},\hat{B}}(\omega)
= G_{
\hat{A},\hat{B}} (\omega+\mathrm{i}0^+)-G_{\hat{A},\hat{B}}
(\omega-\mathrm{i}0^+)$,
where $G_{\hat{A},\hat{B}}(z)$ denotes the zero-temperature Green's
function.  Here we will take $\hat{B}=\hat{A}^\dagger$, yielding
\begin{align}
G_{\hat{A},\hat{A}^\dagger}(z)&=:\,
G^{(1)}_{\hat{A},\hat{A}^{\dagger}}(z)-s\cdot
G^{(2)}_{\hat{A}^\dagger,\hat{A}}(z)\label{eq:GF_def}\\
G^{(1/2)}_{\hat{X},\hat{Y}}(z)&=\langle
\psi_0|\hat{X}\frac{1}{z\mp(\hat{H}-E_0)}\hat{Y}
|\psi_0\rangle \, , \notag
\end{align}
where $|\psi_0\rangle$ is the ground state of Hamiltonian $\hat{H}$,
$E_0$ is the ground-state energy, and $s=+1/-1$ when
$\hat{A}$ is a bosonic/fermionic operator. One of
several ways  to represent the resolvent in Eq.\
(\ref{eq:GF_def}) is the CFE\cite{Gagliano1987,Hallberg1995}
\begin{align}
 G^{(1)}_{\hat{A},\hat{A}^\dagger}(z)&=
\frac{\langle\psi_0 |\hat{A}\hat{A}^\dagger|\psi_0\rangle}{z-a_0-\frac{b^2_1}{
z-a_1-\frac{b^2_2}{z-..}}}\label{eq:cont_frac}
\, , 
\end{align}
with a similar expression for $G^{(2)}_{\hat{A}\hat{A}^\dagger}$.
The coefficients $a_i,b_i$ of this CFE can be
calculated using the recursion formula 
\begin{align}
|f_0\rangle &=
\hat{A}^\dagger|\psi_0\rangle,\quad|f_{n+1}
\rangle=H|f_n\rangle-a_n|f_n\rangle-b^2_n|f_{n-1}\rangle
\notag\\
a_n&=\langle f_n|H|f_n\rangle/\langle f_n|f_n\rangle, \notag\\
b_n&=\langle f_n|f_n\rangle/\langle f_{n-1}|f_{n-1}\rangle,\qquad
b_0=0 \, ,\label{recursion}
\end{align}
which is essentially the one used in the Lanczos method,
but with starting vector $|f_0\rangle$ replacing a random vector.
Thus, we will call the $|f_i\rangle$ Lanczos vectors in
the following.
Hallberg\cite{Hallberg1995} used the CFE (\ref{eq:cont_frac}) and the
recursion formulae (\ref{recursion}) to obtain spectral functions via a
multi-target DMRG, i.e., by optimizing the
DMRG basis for the ground state and the Lanczos states simultaneously.


The structure of the recursion formula (\ref{recursion})
suggests implementing an iterative method to calculate the Lanczos
vectors in the DMRG that optimizes the basis for only the three
Lanczos vectors needed at each recursion step. 
One initially calculates the ground state
$| \psi_0\rangle$ of the system to the desired accuracy with the usual
DMRG algorithm. 
One then performs single finite-system sweeps, simultaneously
targeting the additional vectors $|f_0\rangle$ and $|f_1\rangle$, from which one
can evaluate $a_0$, $a_1$, and $b_1$. The recursion
proceeds by replacing $|\psi_0\rangle$, $|f_0\rangle$, and $|f_1\rangle$ by
$|f_0\rangle$, $|f_1\rangle$, and the new vector $|f_2\rangle$. 
At this point, a technical subtlety arises: 
$|f_0\rangle$ and $|f_1\rangle$ cannot be recalculated because
there is no longer a condition to optimize them.
Instead, we transform the wave function from the previous
finite-size DMRG step to the new superblock
configuration \cite{White1996} at every step of the DMRG sweep. 
One DMRG sweep through the system thus suffices to calculate the Lanczos vector
$|f_i\rangle$ and the parameters $a_i$ and $b_i$. 
As one needs to target only
three Lanczos vectors simultaneously, the number $m$ of basis states necessary
to obtain an accurate representation
is generally substantially smaller than in the original
algorithm.\cite{Hallberg1995} We will come back to this point later. 
Here we emphasize only that we avoid calculating the ground state at each
DMRG step, speeding up the calculation dramatically;
the most time-consuming part left in the Lanczos iteration is now the
diagonalization of the reduced
density matrix needed to initialize the next DMRG step. After iterating the
recursion relation often enough, one obtains a sequence
$\{(a_i,b_i)\}$ from which one can calculate the Green's function, 
Eq.\ (\ref{eq:cont_frac}). 

\begin{figure}[tb]
\centering
\includegraphics[width=0.91\columnwidth]{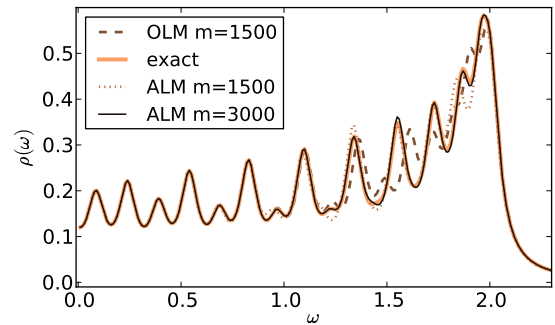}
\caption{\label{fig1} (Color online) Broadened spectral function
for spinless
fermions
($L=40,U=0, \eta=0.05, 200 \text{ Lanczos vectors}$)
calculated with the new adaptive implementation (ALM), the original
implementation (OLM) and compared to the exact solution. The DMRG truncation
number is given by $m$.}
\end{figure}

We test the method on a model of
spinless fermions on a chain with Hamiltonian 
\begin{align}
\hat{H}=&-\sum_i\left( c^\dagger_{i+1} c^{\phantom\dagger}_i +c^\dagger_{i}
c^{\phantom\dagger}_{i+1} \right)+U\sum_i
n_in_{i+1}\, , \notag
\end{align}
where $c^{(\dagger)}_{i}$ denotes the usual creation (annihilation)
operators for an electron at site $i$, and $n_i$ denotes the occupation number
operator. Here we take $A=\hat{c}_i$ and calculate the spectral
function $\rho(x=i,\omega)$ at the center of the chain for a system at
half filling. In Fig.\ \ref{fig1},
we compare the original (OLM) and the ALM implementation of
the CFE with the exact solution on a chain of length
$L=40$ in the free-fermion limit, $U=0$. Both implementations do not give a
clear hint as to when to stop the Lanczos iteration.
We have evaluated the broadened spectral functions after
200 Lanczos iterations, see below.
Following K\"uhner and White,\cite{Kuhner1999} we do
not target all Lanczos vectors in the OLM but only a
small fraction of them. We then calculate the remaining vectors
by straightforward application of the recursion. To achieve good convergence of
the DMRG, we  assign a weight of $0.5$ to the ground state and
equal weights to the other Lanczos vectors (one could also give
them weights according to their spectral weights). This large freedom, 
especially in choosing the number of targeted Lanczos
states, is in our a view a disadvantage of the OLM, as the optimal
parameters are not obvious, and the results depend on their choice. Employing the
truncation error as measure of the quality of the spectral function is also not
possible, as it
depends only on the number of targeted states and their weights. 

In Fig.\ \ref{fig1} we have used $m=1500$, $25$ sweeps, and eleven
target states, yielding a maximum truncation error of $10^{-6}$. There is no way
to determine the quality of the representation of the following Lanczos states
(one can only check
their orthogonality), but if we would have tried to target all Lanczos states,
the truncated weight would have been be much higher than $10^{-6}$. 
In addition, the convergence is unstable, as
small changes in the first Lanczos vectors will lead to bigger changes
in subsequent Lanczos vectors. 

In the ALM, we can increase the number of states during the Lanczos
iteration, as we do not need
to calculate the ground state at each step. We notice that the number of
states needed for a good representation of Lanczos vectors increases with the
number of iterations desired.  
For $m=1500$, the maximum truncation error per sweep for the ALM increases
from $10^{-15}$ (ground state and first two Lanczos vectors) to
$10^{-4}$ for the last three Lanczos vectors. As already noted
for the OLM,\cite{Kuhner1999} the low-energy portion
($\omega\lesssim 1$) is well-reproduced by both methods. Deviations occur in the
high-energy part of the spectrum, notably in both position and
weight distribution for the OLM. Here the ALM already shows
much better agreement with the exact solution for $m=1500$. 
Increasing to $m=3000$ for the Lanczos iterations in the ALM, one
almost perfectly reproduces the exact solution. Note that $m=1500$ is
the maximum number of states accessible in the OLM with our computational
resources; the simultaneous optimization of the ground state and all other
Lanczos states prevents calculations with larger $m$.

Another advantage of the ALM is its shorter run-time.
For $m=1500$ our calculations
used  $\approx 30h$ on a standard workstation for the ALM and $\approx
90h$ for the OLM. We
emphasize, however, that the run-time
is determined by several parameters and thus should be interpreted
with some care.  For example, the ALM
scales linearly with the number of Lanczos iterations, whereas the original
implementation is roughly independent of the iteration number. 
The run-time of the latter, however,
strongly depends on the number of  Lanczos states targeted,
because more target states increase the number of DMRG sweeps needed 
to achieve convergence.

\begin{figure}[t!]
\centering
\includegraphics[width=0.91\columnwidth]{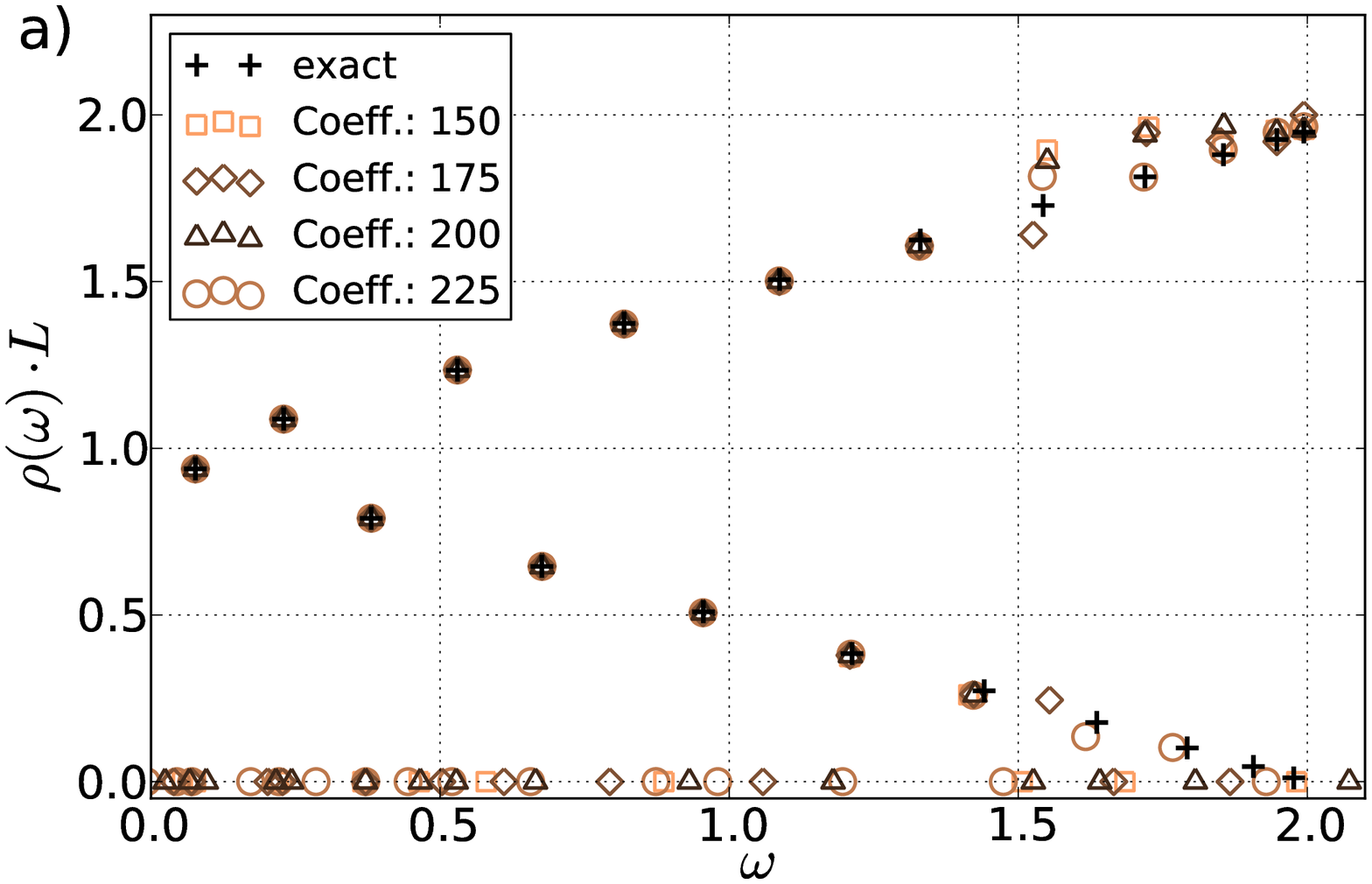}
\includegraphics[width=0.91\columnwidth]{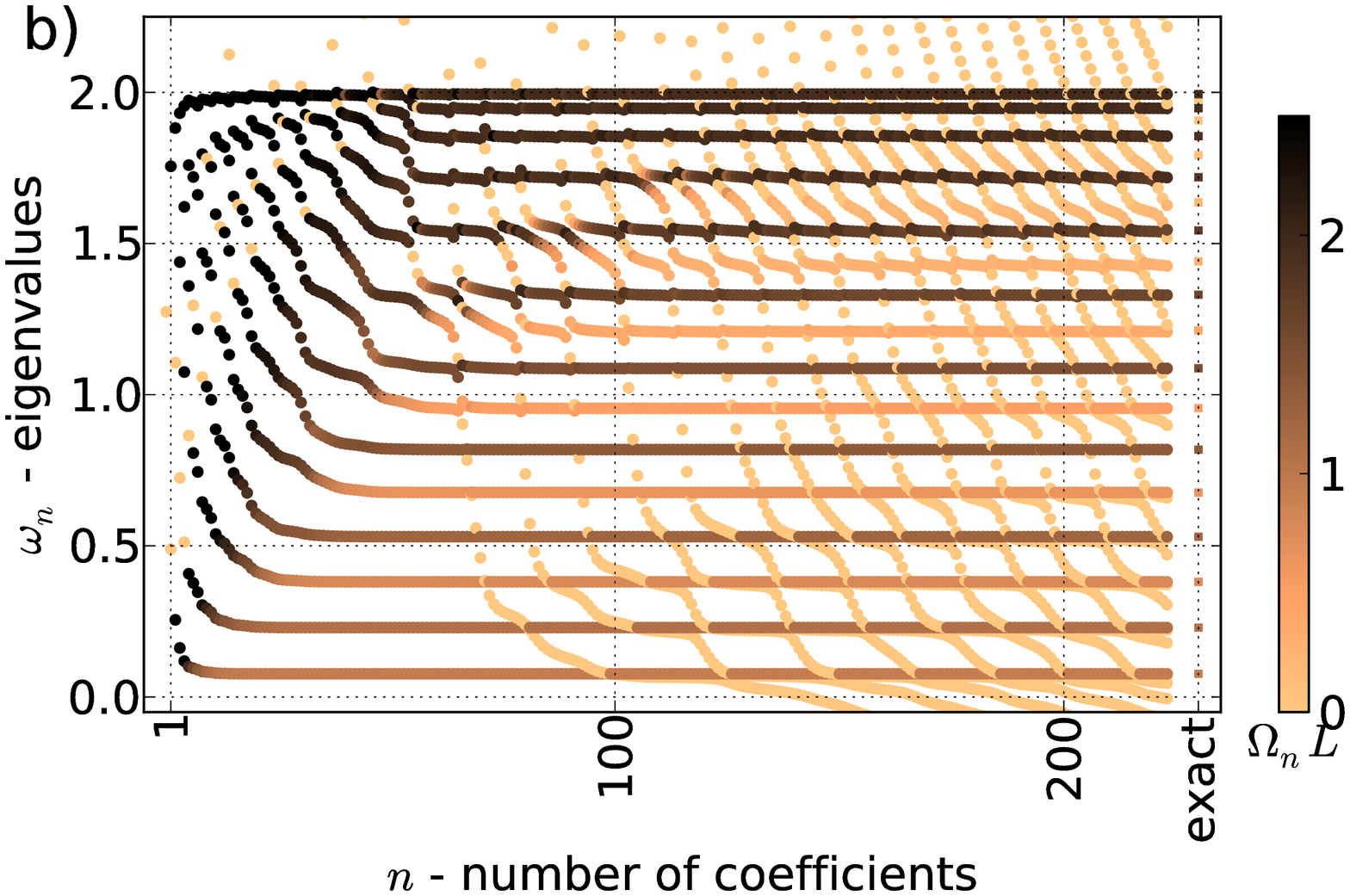}
\caption{\label{fig2-3} (Color online)
Spectral weights evaluated for different numbers of coefficients (a)
and flow of the eigenvalues as
a function of the number of Lanczos steps $i$
(b)
for spinless fermions ($L=40$, $U=0$, $m=3000$).
}
\end{figure} 

Since the CFE gives an analytical expression for the
spectral function, we can take the limit
$\eta\to0$ so that the spectral function
 $\rho_{\hat{A}\hat{A}^\dagger}(\omega) =\,\sum_n \Omega_n^-
\,\delta(\omega-\omega_n) -s \sum_n \Omega_n^+ \,\delta(\omega+\omega_n) $ is a
series of $\delta$-functions, where
$\omega_n=E_n-E_0$, $\Omega_n^{-(+)}=\left|\langle
\psi_0|\hat{A}^{(\dagger)} |n\rangle\right|^2$
and $|n\rangle$ are the eigenstates of $\hat{H}$ with eigenenergies $E_n$.
Evidently, all of the information on the spectral function is contained in
the weights $\Omega_n^\pm$ and poles $\omega_n=\pm\Delta E_n$. Furthermore, one
will always obtain a discrete and finite set of poles for any finite system.
Since the DMRG treats finite systems, the spectral weights and pole positions
can be calculated directly.
Such a calculation has two important advantages:
{\em (i)}
One can study the size-dependence of spectral properties in a very
controlled manner.  
{\em (ii)}
It is possible to obtain precise values for the energies of the low-lying
excitations;\cite{Schneider2008} they are given by the pole positions
$\omega_n$, which can be directly obtained as the eigenvalues of the
tridiagonal matrix
$T_{ij}=a_{i-1}\delta_{ij}+b_i\,\left(\delta_{j,i+1}+\delta_{i+1,j}\right)$.\cite{Kuhner1999} 
The poles of the Green's function are located on the real axis and they are
discrete. Therefore, one can integrate along a closed path ${\cal C}$ in the
complex plane chosen to
enclose one single singularity, say at $\omega_n$. According to the residue
theorem, the weight $\Omega_n$ of this pole is then
\begin{align}
2\pi \mathrm{i}~\Omega_n 
&= \int^{\infty}_{-\infty} \left[G(\omega_n-\epsilon+\mathrm{i} \gamma) 
 -  G(\omega_n+\epsilon+\mathrm{i}\gamma)\right]\,d\gamma \notag
\, .
\end{align}
The parameter $\epsilon$ must be chosen to be smaller than the distance to the
next eigenvalue, $\epsilon <|\omega_n-\omega_{n\pm1}|$. 
Using this procedure with
the position of the poles known, the spectral weights can be calculated to high
precision using numerical integration. 

In Fig.\ \ref{fig2-3}(a) we compare the weights and
positions of the poles calculated with the DMRG and the ALM to the exact values
on a chain of length $L=40$, which are given by 
$\Omega_n = \frac{2}{L+1}|\sin((k_f+k_n)x)|^2$, $\omega_n=2\cos(k_n)$,
where $k_n=\frac{\pi n}{L+1}$, $n\in\{1,\ldots,L\}$, and $k_f$ is the Fermi wave
vector. For small energies,
the agreement is nearly perfect, while deviations
occur for large energies. The quality of the agreement depends on the number of
coefficients $a_i$, $b_i$ taken into account.
One finds that after an
initial improvement, no improvement occurs when one further
increases the number of coefficients.

The origin of this behavior can be understood from Fig.\ \ref{fig2-3}(b), which
depicts the flow of the eigenvalues of the
matrix $T_{ij}$ as a function of the number of Lanczos coefficients.
The convergence of the first few eigenvalues is evidently rapid. The
flow of the eigenvalues therefore gives a nice criterion 
for stopping the iterations (also for the OLM).
However, after
approximately $50$ iterations, an eigenvalue with nearly vanishing weight
appears and subsequently moves rapidly to zero energy.
More such eigenvalues follow, at rapidly increasing frequency. This
appearance of so-called ``ghost'' eigenvalues is well-known 
in the Lanczos method. These ghost
eigenvalues are caused by the loss of orthogonality of the Lanczos vectors
due to numerical error.\cite{Cullum1985,Bai2000} Their appearance here is
therefore not
surprising, in particular because the calculated Lanczos vectors also include an
error from the approximate DMRG representation. We find that this effect is
enhanced as one reduces the number of states within the DMRG. 
Thus, the study of the flow of the eigenvalues makes it possible 
to control the quality of the spectral function.  
Note that such ghost eigenvalues also occur
within the OLM. As far as we know, this problem was never addressed in
detail for this method. While the ghost
eigenvalues seem, at first glance, to be a serious problem,
we emphasize that, for the spectral function in particular, they do not appear to
cause real
harm because they possess only very small weight. This is evident from the
scaling in Fig.\ \ref{fig2-3}(b) and also from the fact that all the ghost
eigenvalues are located on the abscissa in Fig.\
\ref{fig2-3}(a).
This observation is just an empirical one at present. 
However, as long as this remains true,
the method will be insensitive with respect to the
occurrence of ghost eigenvalues, except for regions with very small spectral
weight. Here it is difficult to distinguish
ghosts from small but real spectral weights. 
One possibility is to examine their convergence.\cite{Cullum1985,Bai2000} 
Another problem with ghosts
is that they lead to a violation of sum rules due to double-counting. 
If one adds
up just the real spectral values there will be a missing weight that is
given by the sum of all ghost values. As long as the weight of the ghost values
is small, this will not lead to severe violations of sum rules and can
be used as a measure for the ghost problem. \footnote{Another promising way to
deal with this problem would be to
carry out  a partial or complete reorthogonalization of the calculated Lanczos
 vectors. This can be done by transforming each of the previous Lanczos
 vectors to the new DMRG basis (without adding them to the density
 matrix). This should reduce the ghost problem, but will increase
 computing time and memory requirements.}
Another, in our opinion, much more severe
problem is the poor convergence at large
energies.
This could be addressed by
spectral transformations, which are a standard method for improving
convergence of excited states within the Lanczos method.\cite{Bai2000}

\begin{figure}[tb]
\centering
\includegraphics[width=0.91\columnwidth]{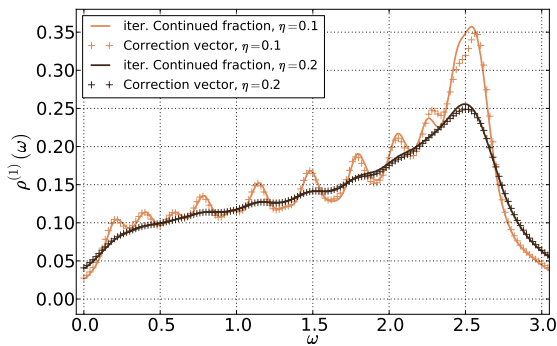}
\caption{\label{fig5} (Color online) Comparison of the spectral weight
calculated using the adaptive Lanczos-vector method with that calculated using
the correction vector method for different broadenings $\eta$
($U=1.0,m_{CV}=300,m_{ALM}=3000,250 \text{ Lanczos vectors} $).}
\end{figure}

We now consider finite interaction, $U=1$, and compare our method to the
correction vector method.\cite{Ramasesha1996,Kuhner1999} The correction
vector can be efficiently
calculated by minimizing a functional at every step of the
DMRG,\cite{Jeckelmann2002} but one must include a finite broadening in this
process. Each broadening requires a separate DMRG run, in contrast to the ALM,
which can evaluate the Green's function readily for any set of frequencies and
arbitrary broadening once the Lanczos coefficients have been calculated. In
Fig.\ \ref{fig5}, we compare results from the two methods for two
different values of the broadening,  $\eta=0.1$ and $0.2$. We find very good
agreement overall, especially for small energies. As expected, there are
deviations at high energies, which become more pronounced with decreasing
broadening.


We have shown that the ALM is capable of calculating the spectral weights and
poles accurately within the DMRG. In
contrast to the OLM, we obtain the correct
weights and poles of a Green's function up to energies of the order of half the
bandwidth with only moderate resources, and, as long as the spectral
weight is not too small, a reasonable reproduction of the spectrum even at
larger energies. A clear advantage of the
CFE in general is that it
is based on an analytical expression for the Green's function, making it
possible to evaluate it at an arbitrary set of frequencies with any possible
broadening without having to repeat the DMRG calculation. In addition, one can
extract irreducible quantities such as self-energies directly as continued
fractions. The inaccuracies that appear at higher energies can be traced to a
loss of orthogonality of the Lanczos vectors due to systematic and numerical
errors. 
In order to improve the accuracy of spectral functions at higher energies, the
truncation number $m$ must be increased. This increase in $m$ can become quite 
resource-intensive in standard DMRG implementations. However, an improvement in
efficiency could be achieved by using the Matrix-Product-State (MPS)
formulation of the DMRG.\cite{Ostlund1995,*Dukelsky1998,*Takasaki1999}
Within the MPS formulation, one can variationally optimize the three Lanczos
vectors in the recursion formula separately, potentially leading to much better
performance. Another advantage is that, within the MPS formulation, the
Hamiltonian can be represented exactly (see Ref.\ \onlinecite{Schollwoeck2011}
for an introduction), thereby further reducing systematic errors. Finally, we
point out that this recursive approach to calculate a special basis
for the evaluation of spectral functions is not limited to the Lanczos basis. 
For example, expansion in Chebyshev polynomials leads to a similar
recursion formula.\cite{Weisse2006}


This work was supported by the DFG via
SFB 602 and grant HO~2325/4-2. R.P.\ is supported by JSPS and the AvH.


\textit{Note added:} After completion of this work, we learned that
Holzner \textit{et al.}\cite{Holzner2011} have shown
that an adaptive method using Chebyshev
polynomials in combination with MPS is highly efficient and gives accurate 
spectral functions.

\bibliography{Literatur}

\begin{thebibliography}{24}%
\makeatletter
\providecommand \@ifxundefined [1]{%
 \@ifx{#1\undefined}
}%
\providecommand \@ifnum [1]{%
 \ifnum #1\expandafter \@firstoftwo
 \else \expandafter \@secondoftwo
 \fi
}%
\providecommand \@ifx [1]{%
 \ifx #1\expandafter \@firstoftwo
 \else \expandafter \@secondoftwo
 \fi
}%
\providecommand \natexlab [1]{#1}%
\providecommand \enquote  [1]{``#1''}%
\providecommand \bibnamefont  [1]{#1}%
\providecommand \bibfnamefont [1]{#1}%
\providecommand \citenamefont [1]{#1}%
\providecommand \href@noop [0]{\@secondoftwo}%
\providecommand \href [0]{\begingroup \@sanitize@url \@href}%
\providecommand \@href[1]{\@@startlink{#1}\@@href}%
\providecommand \@@href[1]{\endgroup#1\@@endlink}%
\providecommand \@sanitize@url [0]{\catcode `\\12\catcode `\$12\catcode
  `\&12\catcode `\#12\catcode `\^12\catcode `\_12\catcode `\%12\relax}%
\providecommand \@@startlink[1]{}%
\providecommand \@@endlink[0]{}%
\providecommand \url  [0]{\begingroup\@sanitize@url \@url }%
\providecommand \@url [1]{\endgroup\@href {#1}{\urlprefix }}%
\providecommand \urlprefix  [0]{URL }%
\providecommand \Eprint [0]{\href }%
\providecommand \doibase [0]{http://dx.doi.org/}%
\providecommand \selectlanguage [0]{\@gobble}%
\providecommand \bibinfo  [0]{\@secondoftwo}%
\providecommand \bibfield  [0]{\@secondoftwo}%
\providecommand \translation [1]{[#1]}%
\providecommand \BibitemOpen [0]{}%
\providecommand \bibitemStop [0]{}%
\providecommand \bibitemNoStop [0]{.\EOS\space}%
\providecommand \EOS [0]{\spacefactor3000\relax}%
\providecommand \BibitemShut  [1]{\csname bibitem#1\endcsname}%
\let\auto@bib@innerbib\@empty
\bibitem [{\citenamefont {White}(1992)}]{White1992}%
  \BibitemOpen
  \bibfield  {author} {\bibinfo {author} {\bibfnamefont {S.~R.}\ \bibnamefont
  {White}},\ }\href {\doibase 10.1103/PhysRevLett.69.2863} {\bibfield
  {journal} {\bibinfo  {journal} {Phys. Rev. Lett.}\ }\textbf {\bibinfo
  {volume} {69}},\ \bibinfo {pages} {2863} (\bibinfo {year}
  {1992})}\BibitemShut {NoStop}%
\bibitem [{\citenamefont {White}(1993)}]{White1993}%
  \BibitemOpen
  \bibfield  {author} {\bibinfo {author} {\bibfnamefont {S.~R.}\ \bibnamefont
  {White}},\ }\href {\doibase 10.1103/PhysRevB.48.10345} {\bibfield  {journal}
  {\bibinfo  {journal} {Phys. Rev. B}\ }\textbf {\bibinfo {volume} {48}},\
  \bibinfo {pages} {10345} (\bibinfo {year} {1993})}\BibitemShut {NoStop}%
\bibitem [{\citenamefont {Hallberg}(1995)}]{Hallberg1995}%
  \BibitemOpen
  \bibfield  {author} {\bibinfo {author} {\bibfnamefont {K.~A.}\ \bibnamefont
  {Hallberg}},\ }\href {\doibase 10.1103/PhysRevB.52.R9827} {\bibfield
  {journal} {\bibinfo  {journal} {Phys. Rev. B}\ }\textbf {\bibinfo {volume}
  {52}},\ \bibinfo {pages} {R9827} (\bibinfo {year} {1995})}\BibitemShut
  {NoStop}%
\bibitem [{\citenamefont {K\"uhner}\ and\ \citenamefont
  {White}(1999)}]{Kuhner1999}%
  \BibitemOpen
  \bibfield  {author} {\bibinfo {author} {\bibfnamefont {T.~D.}\ \bibnamefont
  {K\"uhner}}\ and\ \bibinfo {author} {\bibfnamefont {S.~R.}\ \bibnamefont
  {White}},\ }\href {\doibase 10.1103/PhysRevB.60.335} {\bibfield  {journal}
  {\bibinfo  {journal} {Phys. Rev. B}\ }\textbf {\bibinfo {volume} {60}},\
  \bibinfo {pages} {335} (\bibinfo {year} {1999})}\BibitemShut {NoStop}%
\bibitem [{\citenamefont {Ramasesha}\ \emph {et~al.}(1996)\citenamefont
  {Ramasesha}, \citenamefont {Pati}, \citenamefont {Krishnamurthy},
  \citenamefont {Shuai},\ and\ \citenamefont {Br\'edas}}]{Ramasesha1996}%
  \BibitemOpen
  \bibfield  {author} {\bibinfo {author} {\bibfnamefont {S.}~\bibnamefont
  {Ramasesha}}, \bibinfo {author} {\bibfnamefont {S.~K.}\ \bibnamefont {Pati}},
  \bibinfo {author} {\bibfnamefont {H.~R.}\ \bibnamefont {Krishnamurthy}},
  \bibinfo {author} {\bibfnamefont {Z.}~\bibnamefont {Shuai}}, \ and\ \bibinfo
  {author} {\bibfnamefont {J.~L.}\ \bibnamefont {Br\'edas}},\ }\href {\doibase
  10.1103/PhysRevB.54.7598} {\bibfield  {journal} {\bibinfo  {journal} {Phys.
  Rev. B}\ }\textbf {\bibinfo {volume} {54}},\ \bibinfo {pages} {7598}
  (\bibinfo {year} {1996})}\BibitemShut {NoStop}%
\bibitem [{\citenamefont {Benthien}\ \emph {et~al.}(2004)\citenamefont
  {Benthien}, \citenamefont {Gebhard},\ and\ \citenamefont
  {Jeckelmann}}]{Benthien2004}%
  \BibitemOpen
  \bibfield  {author} {\bibinfo {author} {\bibfnamefont {H.}~\bibnamefont
  {Benthien}}, \bibinfo {author} {\bibfnamefont {F.}~\bibnamefont {Gebhard}}, \
  and\ \bibinfo {author} {\bibfnamefont {E.}~\bibnamefont {Jeckelmann}},\
  }\href {\doibase 10.1103/PhysRevLett.92.256401} {\bibfield  {journal}
  {\bibinfo  {journal} {Phys. Rev. Lett.}\ }\textbf {\bibinfo {volume} {92}},\
  \bibinfo {pages} {256401} (\bibinfo {year} {2004})}\BibitemShut {NoStop}%
\bibitem [{\citenamefont {Shirakawa}\ and\ \citenamefont
  {Jeckelmann}(2009)}]{Shirakawa2009}%
  \BibitemOpen
  \bibfield  {author} {\bibinfo {author} {\bibfnamefont {T.}~\bibnamefont
  {Shirakawa}}\ and\ \bibinfo {author} {\bibfnamefont {E.}~\bibnamefont
  {Jeckelmann}},\ }\href {\doibase 10.1103/PhysRevB.79.195121} {\bibfield
  {journal} {\bibinfo  {journal} {Phys. Rev. B}\ }\textbf {\bibinfo {volume}
  {79}},\ \bibinfo {pages} {195121} (\bibinfo {year} {2009})}\BibitemShut
  {NoStop}%
\bibitem [{\citenamefont {Nishimoto}\ and\ \citenamefont
  {Jeckelmann}(2004)}]{Nishimoto2004}%
  \BibitemOpen
  \bibfield  {author} {\bibinfo {author} {\bibfnamefont {S.}~\bibnamefont
  {Nishimoto}}\ and\ \bibinfo {author} {\bibfnamefont {E.}~\bibnamefont
  {Jeckelmann}},\ }\href@noop {} {\bibfield  {journal} {\bibinfo  {journal} {J.
  Phys.: Condens. Matter}\ }\textbf {\bibinfo {volume} {16}},\ \bibinfo {pages}
  {613} (\bibinfo {year} {2004})}\BibitemShut {NoStop}%
\bibitem [{\citenamefont {Weichselbaum}\ \emph {et~al.}(2009)\citenamefont
  {Weichselbaum}, \citenamefont {Verstraete}, \citenamefont {Schollw\"ock},
  \citenamefont {Cirac},\ and\ \citenamefont {von Delft}}]{Weichselbaum2009}%
  \BibitemOpen
  \bibfield  {author} {\bibinfo {author} {\bibfnamefont {A.}~\bibnamefont
  {Weichselbaum}}, \bibinfo {author} {\bibfnamefont {F.}~\bibnamefont
  {Verstraete}}, \bibinfo {author} {\bibfnamefont {U.}~\bibnamefont
  {Schollw\"ock}}, \bibinfo {author} {\bibfnamefont {J.~I.}\ \bibnamefont
  {Cirac}}, \ and\ \bibinfo {author} {\bibfnamefont {J.}~\bibnamefont {von
  Delft}},\ }\href {\doibase 10.1103/PhysRevB.80.165117} {\bibfield  {journal}
  {\bibinfo  {journal} {Phys. Rev. B}\ }\textbf {\bibinfo {volume} {80}},\
  \bibinfo {pages} {165117} (\bibinfo {year} {2009})}\BibitemShut {NoStop}%
\bibitem [{\citenamefont {Raas}\ and\ \citenamefont {Uhrig}(2005)}]{Raas2005}%
  \BibitemOpen
  \bibfield  {author} {\bibinfo {author} {\bibfnamefont {C.}~\bibnamefont
  {Raas}}\ and\ \bibinfo {author} {\bibfnamefont {G.~S.}\ \bibnamefont
  {Uhrig}},\ }\href {\doibase 10.1140/epjb/e2005-00194-3} {\bibfield  {journal}
  {\bibinfo  {journal} {Eur. Phys. J. B}\ }\textbf {\bibinfo {volume} {45}},\
  \bibinfo {pages} {293} (\bibinfo {year} {2005})}\BibitemShut {NoStop}%
\bibitem [{\citenamefont {Pereira}\ \emph {et~al.}(2009)\citenamefont
  {Pereira}, \citenamefont {White},\ and\ \citenamefont
  {Affleck}}]{Pereira2009}%
  \BibitemOpen
  \bibfield  {author} {\bibinfo {author} {\bibfnamefont {R.~G.}\ \bibnamefont
  {Pereira}}, \bibinfo {author} {\bibfnamefont {S.~R.}\ \bibnamefont {White}},
  \ and\ \bibinfo {author} {\bibfnamefont {I.}~\bibnamefont {Affleck}},\ }\href
  {\doibase 10.1103/PhysRevB.79.165113} {\bibfield  {journal} {\bibinfo
  {journal} {Phys. Rev. B}\ }\textbf {\bibinfo {volume} {79}},\ \bibinfo
  {pages} {165113} (\bibinfo {year} {2009})}\BibitemShut {NoStop}%
\bibitem [{\citenamefont {Gagliano}\ and\ \citenamefont
  {Balseiro}(1987)}]{Gagliano1987}%
  \BibitemOpen
  \bibfield  {author} {\bibinfo {author} {\bibfnamefont {E.~R.}\ \bibnamefont
  {Gagliano}}\ and\ \bibinfo {author} {\bibfnamefont {C.~A.}\ \bibnamefont
  {Balseiro}},\ }\href {\doibase 10.1103/PhysRevLett.59.2999} {\bibfield
  {journal} {\bibinfo  {journal} {Phys. Rev. Lett.}\ }\textbf {\bibinfo
  {volume} {59}},\ \bibinfo {pages} {2999} (\bibinfo {year}
  {1987})}\BibitemShut {NoStop}%
\bibitem [{\citenamefont {White}(1996)}]{White1996}%
  \BibitemOpen
  \bibfield  {author} {\bibinfo {author} {\bibfnamefont {S.~R.}\ \bibnamefont
  {White}},\ }\href {\doibase 10.1103/PhysRevLett.77.3633} {\bibfield
  {journal} {\bibinfo  {journal} {Phys. Rev. Lett.}\ }\textbf {\bibinfo
  {volume} {77}},\ \bibinfo {pages} {3633} (\bibinfo {year}
  {1996})}\BibitemShut {NoStop}%
\bibitem [{\citenamefont {Schneider}\ \emph {et~al.}(2008)\citenamefont
  {Schneider}, \citenamefont {Struck}, \citenamefont {Bortz},\ and\
  \citenamefont {Eggert}}]{Schneider2008}%
  \BibitemOpen
  \bibfield  {author} {\bibinfo {author} {\bibfnamefont {I.}~\bibnamefont
  {Schneider}}, \bibinfo {author} {\bibfnamefont {A.}~\bibnamefont {Struck}},
  \bibinfo {author} {\bibfnamefont {M.}~\bibnamefont {Bortz}}, \ and\ \bibinfo
  {author} {\bibfnamefont {S.}~\bibnamefont {Eggert}},\ }\href {\doibase
  10.1103/PhysRevLett.101.206401} {\bibfield  {journal} {\bibinfo  {journal}
  {Phys. Rev. Lett.}\ }\textbf {\bibinfo {volume} {101}},\ \bibinfo {pages}
  {206401} (\bibinfo {year} {2008})}\BibitemShut {NoStop}%
\bibitem [{\citenamefont {Cullum}\ and\ \citenamefont
  {Willoughby}(1985)}]{Cullum1985}%
  \BibitemOpen
  \bibfield  {author} {\bibinfo {author} {\bibfnamefont {J.}~\bibnamefont
  {Cullum}}\ and\ \bibinfo {author} {\bibfnamefont {R.}~\bibnamefont
  {Willoughby}},\ }\href@noop {} {\emph {\bibinfo {title} {Lanczos algorithms
  for large symmetric eigenvalue computations}}},\ Vol.~\bibinfo {volume} {2}\
  (\bibinfo  {publisher} {Birkhäuser, Boston},\ \bibinfo {year}
  {1985})\BibitemShut {NoStop}%
\bibitem [{\citenamefont {Bai}\ \emph {et~al.}(2000)\citenamefont {Bai},
  \citenamefont {Demmel}, \citenamefont {Dongarra}, \citenamefont {Ruhe},\ and\
  \citenamefont {van~der Vorst}}]{Bai2000}%
  \BibitemOpen
  \bibfield  {author} {\bibinfo {author} {\bibfnamefont {Z.}~\bibnamefont
  {Bai}}, \bibinfo {author} {\bibfnamefont {J.}~\bibnamefont {Demmel}},
  \bibinfo {author} {\bibfnamefont {J.}~\bibnamefont {Dongarra}}, \bibinfo
  {author} {\bibfnamefont {A.}~\bibnamefont {Ruhe}}, \ and\ \bibinfo {author}
  {\bibfnamefont {H.}~\bibnamefont {van~der Vorst}},\ }\href@noop {} {\emph
  {\bibinfo {title} {Templates for the Solution of Eigenvalue Problems: A
  Practical Guide}}}\ (\bibinfo  {publisher} {SIAM, Philadelphia},\ \bibinfo
  {year} {2000})\BibitemShut {NoStop}%
\bibitem [{Note1()}]{Note1}%
  \BibitemOpen
  \bibinfo {note} {Another promising way to deal with this problem would be to
  carry out a partial or complete reorthogonalization of the calculated Lanczos
  vectors. This can be done by transforming each of the previous Lanczos
  vectors to the new DMRG basis (without adding them to the density matrix).
  This should reduce the ghost problem, but will increase computing time and
  memory requirements.}\BibitemShut {Stop}%
\bibitem [{\citenamefont {Jeckelmann}(2002)}]{Jeckelmann2002}%
  \BibitemOpen
  \bibfield  {author} {\bibinfo {author} {\bibfnamefont {E.}~\bibnamefont
  {Jeckelmann}},\ }\href {\doibase 10.1103/PhysRevB.66.045114} {\bibfield
  {journal} {\bibinfo  {journal} {Phys. Rev. B}\ }\textbf {\bibinfo {volume}
  {66}},\ \bibinfo {pages} {045114} (\bibinfo {year} {2002})}\BibitemShut
  {NoStop}%
\bibitem [{\citenamefont {\"Ost\-lund}\ and\ \citenamefont
  {Rommer}(1995)}]{Ostlund1995}%
  \BibitemOpen
  \bibfield  {author} {\bibinfo {author} {\bibfnamefont {S.}~\bibnamefont
  {\"Ost\-lund}}\ and\ \bibinfo {author} {\bibfnamefont {S.}~\bibnamefont
  {Rommer}},\ }\href {\doibase 10.1103/PhysRevLett.75.3537} {\bibfield
  {journal} {\bibinfo  {journal} {Phys. Rev. Lett.}\ }\textbf {\bibinfo
  {volume} {75}},\ \bibinfo {pages} {3537} (\bibinfo {year}
  {1995})}\BibitemShut {NoStop}%
\bibitem [{\citenamefont {Dukelsky}\ \emph {et~al.}(1998)\citenamefont
  {Dukelsky}, \citenamefont {Mart\'{\i}n-Delgado}, \citenamefont {Nishino},\
  and\ \citenamefont {Sierra}}]{Dukelsky1998}%
  \BibitemOpen
  \bibfield  {author} {\bibinfo {author} {\bibfnamefont {J.}~\bibnamefont
  {Dukelsky}}, \bibinfo {author} {\bibfnamefont {M.~A.}\ \bibnamefont
  {Mart\'{\i}n-Delgado}}, \bibinfo {author} {\bibfnamefont {T.}~\bibnamefont
  {Nishino}}, \ and\ \bibinfo {author} {\bibfnamefont {G.}~\bibnamefont
  {Sierra}},\ }\href@noop {} {\bibfield  {journal} {\bibinfo  {journal}
  {Europhys. Lett.}\ }\textbf {\bibinfo {volume} {43}},\ \bibinfo {pages} {457}
  (\bibinfo {year} {1998})}\BibitemShut {NoStop}%
\bibitem [{\citenamefont {Takasaki}\ \emph {et~al.}(1999)\citenamefont
  {Takasaki}, \citenamefont {Hikihara},\ and\ \citenamefont
  {Nishino}}]{Takasaki1999}%
  \BibitemOpen
  \bibfield  {author} {\bibinfo {author} {\bibfnamefont {H.}~\bibnamefont
  {Takasaki}}, \bibinfo {author} {\bibfnamefont {T.}~\bibnamefont {Hikihara}},
  \ and\ \bibinfo {author} {\bibfnamefont {T.}~\bibnamefont {Nishino}},\ }\href
  {\doibase 10.1143/JPSJ.68.1537} {\bibfield  {journal} {\bibinfo  {journal}
  {J. Phys. Soc. Jpn.}\ }\textbf {\bibinfo {volume} {68}},\ \bibinfo {pages}
  {1537} (\bibinfo {year} {1999})}\BibitemShut {NoStop}%
\bibitem [{\citenamefont {Schollwöck}(2011)}]{Schollwoeck2011}%
  \BibitemOpen
  \bibfield  {author} {\bibinfo {author} {\bibfnamefont {U.}~\bibnamefont
  {Schollwöck}},\ }\href {\doibase 10.1016/j.aop.2010.09.012} {\bibfield
  {journal} {\bibinfo  {journal} {Ann. Phys.}\ }\textbf {\bibinfo {volume}
  {326}},\ \bibinfo {pages} {96} (\bibinfo {year} {2011})}\BibitemShut
  {NoStop}%
\bibitem [{\citenamefont {Weisse}\ \emph {et~al.}(2006)\citenamefont {Weisse},
  \citenamefont {Wellein}, \citenamefont {Alvermann},\ and\ \citenamefont
  {Fehske}}]{Weisse2006}%
  \BibitemOpen
  \bibfield  {author} {\bibinfo {author} {\bibfnamefont {A.}~\bibnamefont
  {Weisse}}, \bibinfo {author} {\bibfnamefont {G.}~\bibnamefont {Wellein}},
  \bibinfo {author} {\bibfnamefont {A.}~\bibnamefont {Alvermann}}, \ and\
  \bibinfo {author} {\bibfnamefont {H.}~\bibnamefont {Fehske}},\ }\href
  {\doibase 10.1103/RevModPhys.78.275} {\bibfield  {journal} {\bibinfo
  {journal} {Rev. Mod. Phys.}\ }\textbf {\bibinfo {volume} {78}},\ \bibinfo
  {pages} {275} (\bibinfo {year} {2006})}\BibitemShut {NoStop}%
\bibitem [{\citenamefont {Holzner}\ \emph {et~al.}(2011)\citenamefont
  {Holzner}, \citenamefont {Weichselbaum}, \citenamefont {McCulloch},
  \citenamefont {Schollwöck},\ and\ \citenamefont {von Delft}}]{Holzner2011}%
  \BibitemOpen
  \bibfield  {author} {\bibinfo {author} {\bibfnamefont {A.}~\bibnamefont
  {Holzner}}, \bibinfo {author} {\bibfnamefont {A.}~\bibnamefont
  {Weichselbaum}}, \bibinfo {author} {\bibfnamefont {I.~P.}\ \bibnamefont
  {McCulloch}}, \bibinfo {author} {\bibfnamefont {U.}~\bibnamefont
  {Schollwöck}}, \ and\ \bibinfo {author} {\bibfnamefont {J.}~\bibnamefont
  {von Delft}},\ }\href@noop {} {\bibfield  {journal} {\bibinfo  {journal}
  {arXiv:1101.5895}\ } (\bibinfo {year} {2011})}\BibitemShut {NoStop}%
\end{thebibliography}%

\end{document}